\documentclass{aa}
\usepackage{graphics}
\begin{document}

\def\frac{$''$\hspace*{-.1cm}}
\def\deg{$^{\circ}$}
\def\min{$'$}
\def\deg{$^{\circ}$\hspace*{-.1cm}}
\def\min{$'$\hspace*{-.1cm}}
\def\h2{H\,{\sc ii}}
\def\sm{$M_{\odot}$}
\def\lum{$L_{\odot}$}
\def\l{$\lambda$}
\def\ab{$\sim$}
\def\x{$\times$}
\def\sec{s$^{-1}$}
%
%
\def\alii{Al\,{\sc{ii}}}
\def\aliii{Al~{\sc{iii}}}
\def\cii{C\,{\sc{ii}}}
\def\ciii{C\,{\sc{iii}}}
\def\civ{C\,{\sc{iv}}}
\def\caii{Ca\,{\sc{ii}}}
\def\feii{Fe\,{\sc{ii}}}
\def\feiii{Fe\,{\sc{iii}}}
\def\feiv{Fe\,{\sc{iv}}}
\def\fev{Fe\,{\sc{v}}}
\def\ha{H$\alpha$}
\def\hb{H$\beta$}
\def\hi{H\,{\sc{i}}}
\def\hei{He\,{\sc{i}}}
\def\heii{He\,{{\sc ii}}}
\def\nii{N\,{\sc{ii}}}
\def\niii{N\,{\sc{iii}}}
\def\niv{N\,{\sc{iv}}}
\def\nv{N\,{\sc{v}}}
\def\nai{Na~{\sc{i}}}
\def\nei{Ne~{\sc{i}}}
\def\oi{O\,{\sc{i}}}
\def\oiii{O\,{\sc iii}}
\def\oiv{O\,{\sc{iv}}}
\def\ov{O\,{\sc{v}}}
\def\sv{S\,{\sc{v}}}
\def\siii{Si\,{\sc{ii}}}
\def\siiii{Si\,{\sc{iii}}}
\def\siiv{Si\,{\sc{iv}}}

\def\p{P-Cyg}

\title{STIS spectroscopy of newborn massive stars 
   in SMC N81\thanks{Based 
   on observations with 
   the NASA/ESA Hubble Space Telescope obtained at the Space Telescope 
   Science Institute, which is operated by the Association of Universities 
   for Research in Astronomy, Inc., under NASA contract NAS\,5-26555.}}

\offprints{M. Heydari-Malayeri, heydari@obspm.fr}

\date{Received  13 August 2001 / Accepted: 26 October 2001}

\titlerunning{STIS observations of SMC N81}
\authorrunning{Heydari-Malayeri et al.}

\author{M. Heydari-Malayeri\inst{1} 
        \and 
        M.\,R. Rosa \inst{2,}\thanks{Affiliated to the Astrophysics
        Division, Space Science Department of the European Space
        Agency.}
        \and   
        D. Schaerer \inst{3} 
        \and
        F. Martins \inst{3}
        \and
        V. Charmandaris \inst{4}
}        

\institute{{\sc demirm}, Observatoire de Paris, 61 Avenue de l'Observatoire, 
F-75014 Paris, France 
\and
Space Telescope European Coordinating Facility, European Southern
Observatory, Karl-Schwarzschild-Strasse-2, D-85748 Garching bei
M\"unchen, Germany
\and 
Laboratoire d'Astrophysique, Observatoire Midi-Pyr\'en\'ees, 14,
Avenue E. Belin, F-31400 Toulouse, France
\and 
Cornell University, Astronomy Department, 106 Space Sciences Bldg.,
Ithaca, NY 14853, USA
}

\abstract{Using {\it Hubble Space Telescope} observations with
STIS, we study the main exciting stars of  N81, a high excitation
compact \h2\, region in the Small Magellanic Cloud (SMC). These far
UV observations are the first spectroscopic measurements of stars in
such a region and reveal features characteristic of an O6--O8 stellar
type.  The astonishing weakness of their wind profiles and their
sub-luminosity (up to \ab\,2 mag fainter in  $M_{V}$ than the corresponding
dwarfs) make these stars a unique stellar population in the Magellanic
Clouds.  Our analysis suggests that they are probably in the
Hertzsprung-Russell diagram locus of a particularly young class of
massive stars, the so-called Vz luminosity class, as they are arriving
on the zero age main sequence.
}

\maketitle

\section{Introduction}

Understanding the formation of massive stars, which is still a largely
unsolved problem, requires studying them at the earliest phases where
they can be reached through the enshrouding material at different
wavelengths.  While high resolution radio continuum observations allow
the investigation of ultracompact \h2\, regions formed around newborn
massive stars (Churchwell \cite{chur}), high angular resolution
observations in the ultraviolet, visible, and infrared  are  also
necessary to access accurate physical parameters of these stars  
in order to identify their evolutionary states (Walborn \& Fitzpatrick
\cite{wal}, Walborn et al.\ \cite{wal95b}, 
Hanson et al.\ \cite{hanson}).  In particular UV observations are of
prime importance since massive stars emit the bulk of their energy in
this wavelength range.   In practice though observing newborn
massive stars is not straightforward for several reasons. 
Mainly, they are very rare, and the relatively small
evolutionary timescales involved make it difficult to catch them just
at this very point in their evolution, that is when they become
observable in the UV and visible   (Yorke \& Kr\"ugel \cite{yk}, Shu et
al.\ \cite{shu}, Palla \& Stahler \cite{pal}, Beech \& Mitalas
\cite{bee}, Bernasconi \& Maeder \cite{bern}). \\

We have amply argued that the compact \h2\, regions  known as
HEBs (High Excitation Blobs) provide the best opportunities for a
direct access to massive stars at very early stages of their evolution
(Heydari-Malayeri et al.\ \cite{hey01a} and
references therein).  The members of this distinct and very rare
class of ionized nebulae in the Magellanic Clouds are small and
compact ($\sim$\,5\frac\, to 10\frac\, in diameter corresponding to
$\sim$\,1.5--3.0\,pc), in contrast to the typical \h2 regions in those
galaxies, which are extended structures (sizes of several arc minutes
corresponding to more than 50\,pc, powered by a large number of
exciting stars).  In general, HEBs are also heavily affected by local
dust, as one would expect from their very young age (Heydari-Malayeri
et al.\ \cite{hey01a} and references therein, see also Israel \&
Koornneef \cite{ik91}). And their study is pertinent to understanding
the process of massive star formation especially in the context of the
Magellanic Clouds. \\

Our recent high resolution imaging with the {\it Hubble Space
Telescope} (GO\,6563, GO\,8246) using the Wide Field Planetary Camera
(WFPC2)  has for the first time resolved several HEBs which  
had appeared
featureless to ground-based telescopes: SMC N81, N88A, LMC 159-5 (the
Papillon nebula), N83B, and N11A (Heydari-Malayeri et al.\ 1999a,
1999b, 1999c, 2001a, 2001b). The {\it HST} observations uncover the so far
hidden stellar content  as well as the nebular features of these compact
nebulae and display a turbulent environment typical of
newborn massive star formation sites: outstanding emission ridges
created by  shocks and cavities sculpted in the ionized gas by the
powerful winds of massive stars, prominent dust structures protruding
from hot gas. The observations  also bring  to light even more compact
\h2\, blobs, immersed in the HEBs,  harboring newborn, hot
stars. \\

\begin{table*}[t]
\caption[]{SMC N81 stars observed with STIS}
\label{phot}
\begin{flushleft}
\begin{tabular}{cccccccc}   
\hline
Star number & $\alpha$ & $\delta$ & $y$      & $b-y$ & $A_{V}$ &  $M_{V}$  \\
            &  (J2000)  & (J2000) & (F547M)  &       &         &           \\
            &           &         & (mag)    & (mag) & (mag)   & (mag)     \\
\hline
1  & 01:09:13.1 & --73:11:38.3 & 14.38 & --0.10 & 0.22 & --4.84  \\
2  & 01:09:13.0 & --73:11:38.0 & 14.87 & --0.11 & 0.19 & --4.32  \\
3  & 01:09:13.4 & --73:11:38.4 & 16.10 & --0.08 & 0.31 & --3.21  \\
4  & 01:09:12.8 & --73:11:38.3 & 17.41 &  +0.07 & 1.02 & --2.61  \\
5  & 01:09:13.3 & --73:11:37.6 & 18.29 & --0.05 & 0.46 & --1.17\\
8  & 01:09:12.8 & --73:11:40.2 & 17.84 &  +0.15 & 1.40 & --2.56  \\
11 & 01:09:13.7 & --73:11:33.3 & 15.74 & --0.10 & 0.22 & --3.48  \\
13 & 01:09:16.1 & --73:11:29.1 & 16.65 & --0.08 & 0.31 & --2.66  \\
\hline
\end{tabular}
\end{flushleft}
\end{table*}

 The present paper is devoted to N81, also known as DEM\,138 (Henize 
\cite{henize}, Davies et al.\ \cite{dem}), a nebula only
\ab\,10\frac\, across and located in the Shapley Wing
at \ab\,1\deg.2 (\ab\,1.2\,kpc) from the main body of
the SMC. A first detailed study of this compact \h2\, region carried
out by Heydari-Malayeri et al.\ (\cite{hey88}), revealed its nature and
some of its physical characteristics: gas density and temperature,
chemical composition, mass, age, etc.  Subsequently, near infrared
observations showed the presence
of H$_{2}$ emission towards N81 (Israel \& Koornneef \cite{ik88}),
while $^{12}$CO\,(1\.-\.0) emission at two points towards this
\h2\, region was also detected (Israel et al. \cite{is}).
However, due to the lack of sufficient spatial resolution, it was not
possible to view and study the exciting star(s) hidden inside the
ionized gas.  Therefore, the rather important question, 
 which is often challenged by star formation theories, of whether N81
was powered by a single massive star or a cluster of them, remained
unanswered.  This is, however, a critical question for theories of
star formation. \\

High spatial resolution imaging with  {\it HST} allowed us to
resolve N81 and reveal the presence of a tight cluster of newborn
massive stars embedded in this compact nebula (Heydari-Malayeri
et al.\ \cite{hey99}, hereafter Paper I). Six of the stars are grouped
in the core region of $\sim$\,2\frac\, diameter, with a pair of the
main exciting stars in the very center separated by only 0\frac.27 or
0.08 pc.  The images also displayed conspicuous marks of strong stellar
winds, shocks, and ionization fronts characterising  turbulent massive
star forming regions.  Moreover they revealed prominent dust lanes
dividing the nebula into three lobes. One of the lanes
running over 15\frac\, (4.5\,pc) ends in a magnificent curved plume.
A remarkable absorption ``hole'' or dark globule of radius
\ab\,0\frac.25 (\ab\,0.07 pc)  is situated towards the center of 
the \h2\, region, where the extinction reaches higher values
(A$_{V}$\,=\,1.3 mag).  These absorption features are probably parts
of the molecular cloud which has given birth to the massive stars.  \\

From the Str\"omgren {\it uvby} imaging with WFPC2 we carried out the
photometry of some 50 stars towards N81. This allowed us, using
color-magnitude diagrams, to select the main exciting stars of the
region. This paper is devoted to the spectroscopy of these stars.  We
derive spectral classification for these very young massive stars
and study their nature. \\

\section{Observations and reduction}

The General Observer Program No 8246 devoted to observations of N\,81
was performed with Space Telescope Imaging Spectrograph, STIS
(Woodgate et al.\ \cite{wood}) on board {\it HST} on 28 and 31 October
1999.   The spectra were obtained with the far-UV 
Multi-Anode Microchannel Array (MAMA) detector in the G140L mode covering
the wavelength range 1120--1715 \AA. All the observations were made
through the 52\frac\,\x\,0\frac.2 entrance slit. The effective
resolution was 0.6\,\AA\, per pixel of 25 $\mu$m, corresponding to a
dispersion of 24\,\AA\,mm$^{-1}$, or a resolution of 1.2\,\AA\,
(FWHM).  The exposure times were set according to the apparent
magnitudes of the stars in order to equalize the
signal-to-noise ratios (S/N) of the spectrograms.  Total exposure
times varied from 1229 sec (stars \#1 and \#2) to 3169 sec (stars \#3,
\#4, and \#8).  Three relatively faint stars (\#5, \#7, \#10), not
initially scheduled for observations, happened to lie on the slit when
observing their adjacent stars (\#3 and \#11). The S/N ratio is
particularly weak for these stars, yet we present the spectrogram of
star \#5 which shows some interesting features. 
 STIS was also used to
obtain the spectra of the N81 stars in the visible domain.  The
grating G430L covered the range of 2900 to 5700 \AA\, with a
resolution of of 2.73 \AA\, per pixel. The CCD pixels of 21 $\mu$m
yielded a dispersion of 130\,\AA\,mm$^{-1}$. The exposure times ranged
from 24 sec (star \#1) to 750 sec (star \#8).  \\

 The calibrated output products from the standard pipeline use a
default extraction aperture of 22 pixels (0\frac.53 on the sky).  We
carefully reprocessed the 2D images using the most recent calibration
reference files applicable to the observations and extracted the
spectra using slits of both 6 and 2 pixels. We verified the centering
of the stars on the slits, and tested for the effects of different sky
background extraction on our spectra.  The 6 pixel slit yielded
spectra which are very similar to
those produced by the standard pipeline with an insignificant loss in
S/N. Even the 2 pixel slit did not indicate any extraction effect due
to its size other than the expected loss in S/N. Comparing the
resulting line profiles as a function of the aperture width, we found
that other than an increase in noise when going to the smaller slits,
and slightly modifying the slope of the spectra (because of slight
tilt of the spectral images), the line profiles do not change by any
significant amount. The spectra displayed in Figs. 1 and 2 are based
on 6 pixel extraction apertures.

\section{Results}

The N81 stars observed with STIS are listed in Table 1, and  their
physical location can be seen on Fig. 2 of Paper I.  The table also
presents the corresponding photometry of the stars (Paper I).  The
color excesses $E(B-V)$ were derived from $E(b-y)$ using the intrinsic
color $(b-y)_{0}=-0.15$ mag for hot stars (Relyea \& Kurucz \cite{rk})
and assuming that our observed colors represent the standard
Str\"omgren system.  Then the relation $E(B-V)=1.49 E(b-y)$ (Kaltcheva
\& Georgiev
\cite{kg}) was used to transform into the Johnson system, which
finally yielded the extinctions $A_{V}=3.1 E(B-V)$. The estimated 
absolute magnitudes are based on a distance modulus 
$M$\,--\,$m$\,=\,19.0 mag (corresponding to a distance of 63.2\,kpc, e.g. 
Di Benedetto \cite{di} and references therein) and assuming that the 
Str\"omgren $y$ filter is equal to the Johnson $V$. \\

The final reduced spectrograms are presented in Figs. 1 and 2, where
the former figure includes the four brightest stars of the sample,
whereas Fig. 2 displays the fainter ones.  The main stellar features
(\ciii\,\l\,1176, \nv\,\l\,1239, 1243,
\ov\,\l\,1371, \siiv\,\l\l\,1394, 1403, \civ\,\l\,\,1548, 1551, and 
\heii\,\l\,1640)  are distinguished with  tick marks.
The labels appearing below the features indicate cases where
contamination with an interstellar component is possible.  The
identification of \sv\,\l\,1502\,\AA\, is based on the work of Werner
\& Rauch (\cite{wr}). \\

An outstanding aspect of these spectra is the extreme weakness of the
UV wind profiles.   Weak stellar wind features in SMC O stars 
have already been found by several workers (Hatchings \cite{hutchings}, 
Garmany \& Conti \cite{garmany}, 
Walborn et al. \cite{wal95b}), who
ascribed them to the metal deficiency of the SMC leading to a reduced
radiation pressure responsible for driving the winds of early type
stars.  More recent observations have further confirmed this result 
(Smith Neubig \& Bruhweiler \cite{smith_neubig_smc}, Walborn et
al.\ \cite{wal00}).
 However, the wind features observed in the stars of N81
 are even weaker. If we consider the usually stronger wind lines seen
   in O stars (such as the \nv\,\l\,1239, 1243 and \civ\,\l\,1548,
  1551 in dwarfs, or the \siiv\,\l\l\,1394, 1403 in
  giants/supergiants), with the exception perhaps of star \#5, none
  shows the emission part of a P-Cygni profile and the absorption is
  extremely weak particularly for \nv\,\l\l\,1239, 1243.
For stars which we classify as O types (see below), such a behavior is
observed for the first time.

\begin{figure*}
\begin{center}
\resizebox{18cm}{23cm}{\includegraphics{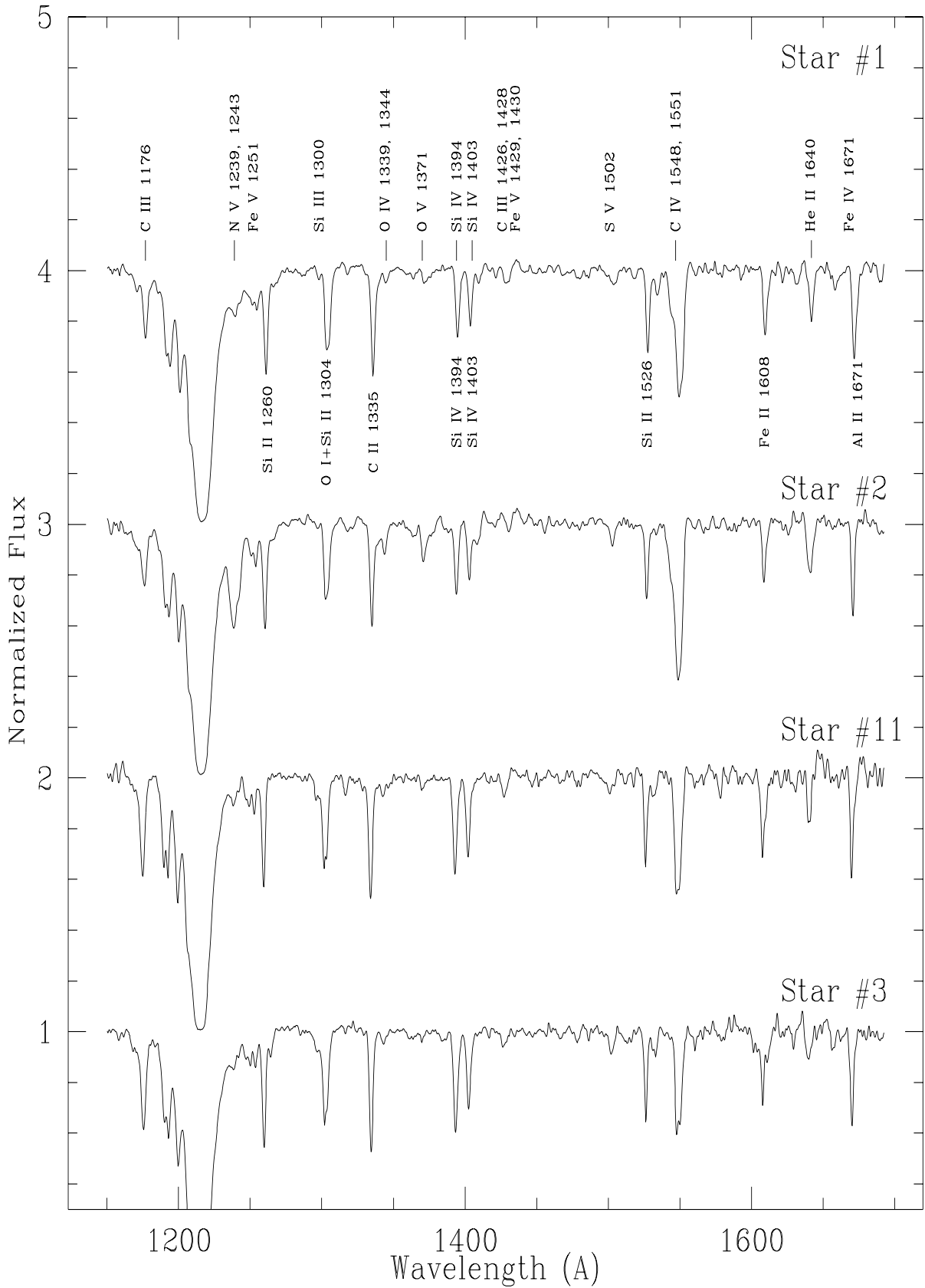}}
\caption{Rectified {\it HST}/STIS ultraviolet spectrograms of the four 
brightest stars in SMC N81. The prominent absorption feature at
\l\,1210\,\AA\, is due to the Ly\,$\alpha$.The wind profiles are
indicated with tick marks and the features possibly contaminated by an
interstellar component are labelled below the lines. }
\end{center}
\end{figure*}

\begin{figure*}
\begin{center}
\resizebox{18cm}{23cm}{\includegraphics{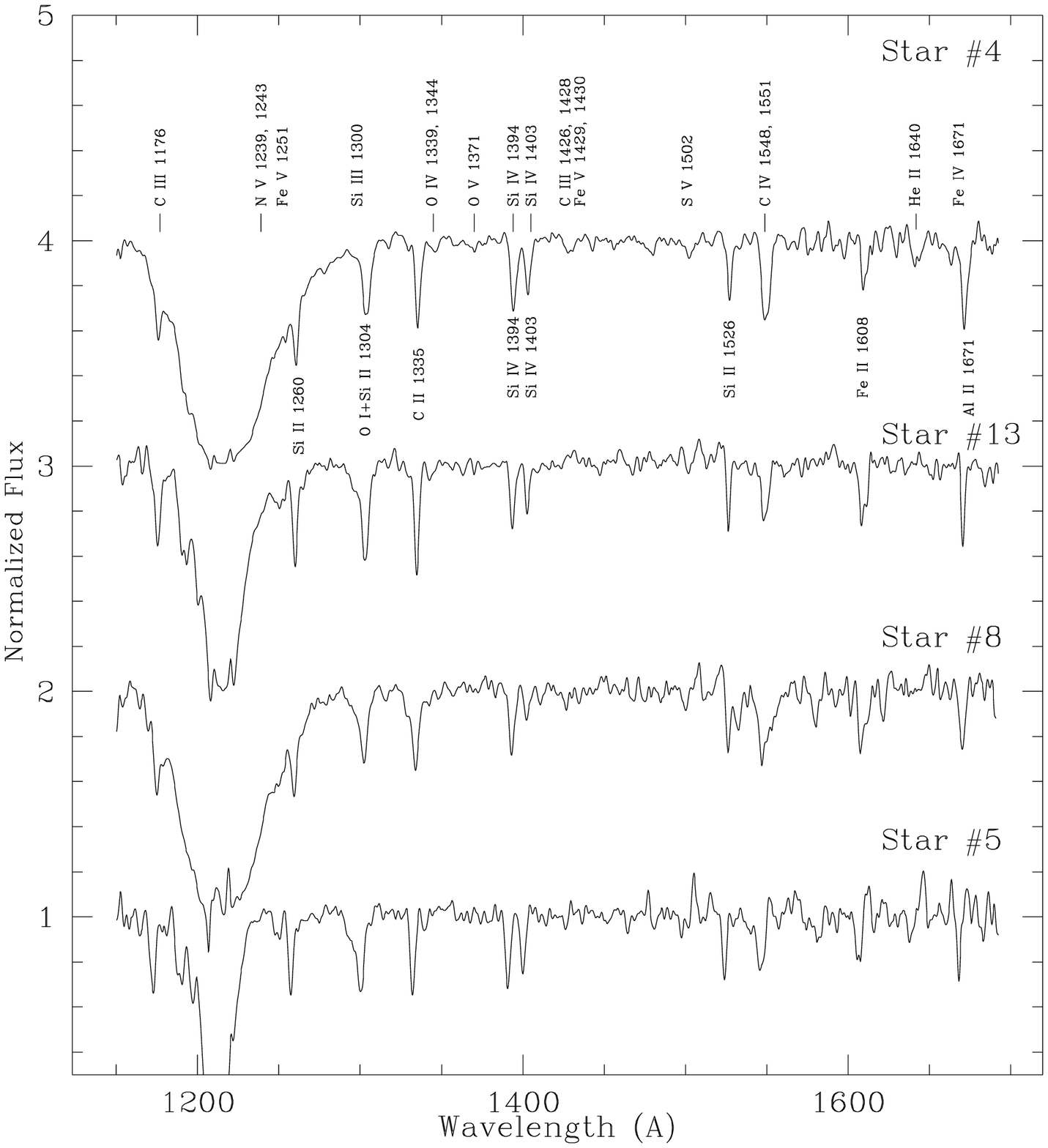}}
\caption{Spectrograms for the remaining  four les bright stars of N81. 
The notation is the same as in Fig. 1.}
\end{center}
\end{figure*}

\subsection{Spectral classification}

Traditionally the spectral classification schemes of stars are based
on  their optical spectra.  However, recent studies comparing spectral
features in the optical and UV have resulted in a global, coherent
picture of classification criteria (Walborn et al.\ \cite{wal85},
\cite{wal95a}, Smith Neubig \& Bruhweiler \cite{smith_neubig_smc},
\cite{smith_neubig_lmc}).  In particular, 
Smith Neubig \& Bruhweiler (\cite{smith_neubig_smc}) 
have proposed a UV classification system for
O and B stars of the SMC which is defined by a set of standard, low
resolution spectra observed with the {\it International Ultraviolet
Explorer (IUE)}.  This UV scheme, which was used by the authors to
derive classifications for 133 O and B stars of the SMC, while
independent of the MK system, shows general agreement with those
deduced from visual data. \\

The low S/N ratio of our STIS optical spectra and in particular
their contamination with strong nebular emission lines limit their 
practical use for spectral
classification. This can be understood  since
N81 is a very bright compact \h2\, region
with strong nebular emission lines  in the visible part of
the spectrum (Paper I).  Although nebular emission lines are
present also in the UV part, they are much less troublesome.  Therefore,
we will use the method put forward by Smith Neubig \& Bruhweiler
(\cite{smith_neubig_smc}).  However, given some morphological
differences of the N81 UV spectra with previously well studied stars,
the limitations of the optical part of the spectrum mentioned earlier,
and the constraints available from the UV classification scheme (Smith
Neubig \& Bruhweiler \cite{smith_neubig_smc}), it is clear that there
is no unique solution to the spectral classification of our
targets. \\

The presence of \heii\,\,\l\,1640 and \ov\,\,\l\,1371 in all spectra,
except perhaps in stars \#13 and \#8 which have lower S/N ratios, is
the first and strongest evidence that we are observing O type stars.
These spectra though display characteristics suggesting that the stars
also belong to the dwarf luminosity class since the features
\siiv\,\l\l\,1394, 1403 as well as \nv\,\l\l\,1239, 1243 and 
\civ\,\l\l\,1548, 1551 are weak (even weaker than usual). 
Star \#5 may be a different case as explained below.
These features are known to increase with luminosity, ranging from
weak \p\, profiles on the main sequence to very pronounced \p\,
profiles in the supergiants (Walborn et al.\ \cite{wal95a}, Smith
Neubig \& Bruhweiler
\cite{smith_neubig_smc}).  A dwarf luminosity class is also supported by the  
optical spectra which show no
\heii\,\l\,\,4686 and \niii\,\l\,4640 emission.  Although the
morphology of the N81 spectra differs qualitatively from that of the
known O types, we may classify them as ``zero age main sequence
dwarf'' Vz, based on the weakness of the wind lines (Walborn \& Parker
\cite{wp}, Walborn \& Blades \cite{wb}, Walborn et al. \cite{wal00}). 
 Note that the original definition of the Vz class is \heii\,\l\,4686
absorption much stronger than \heii\,\l\,4541 or \hei\,\l\,4471, and 
therefore the association of these stars with that class is indirect. \\

No clear distinction between the various subtypes can be made based on
the available spectral features, but it is very likely that all stars
are of a late O type (\ab\,O6\,--\,O8).  This is supported by the
following three facts.  First, the presence of the \ov\,\,\l\,1371
feature indicates a spectral type earlier than $\la$\,O8 (Smith Neubig
\& Bruhweiler \cite{smith_neubig_smc}).  Second, the weakness of the
\siiii\,\,\l\,1300 feature, which appears in the wing of
\oi\,+\,\siii\,\,\l\,1304, excludes much later types.  Finally, the
\nv\,\l\l\,1239, 1243 feature is weaker than \civ\,\l\l\,1548, 1551, 
which is only seen in O types $\geq$\,O6--O7 or alternatively in OC
stars (Walborn \& Panek \cite{wpan}). \\

Star \#5 shows some puzzling emission features in its spectrum
(\l\l\,1480, 1508, 1616, and 1640 \AA), at least one of which, 
\heii\,\l\,1640,  is apparently part of a P-Cyg profile.  It should be
stressed though that the S/N ratio is not high enough to be absolutely
certain about their presence.  If these were indeed wind induced
features, their presence would confirm the fact that the
\civ\,\l\l\,1548, 1551 profile has a particularly marked emission
component compared to the other stars of the sample. Since both of the
\heii\,\l\,1604 and \sv\,\l\,1502 lines show P-Cyg profiles in Of and
Wolf-Rayet spectra (Walborn et al. \cite{wal85}, Willis et
al. \cite{willis}), star \#5 appears to be an Of or WR candidate in
N81.

\begin{figure*}
\begin{center}
\resizebox{15cm}{!}{\includegraphics{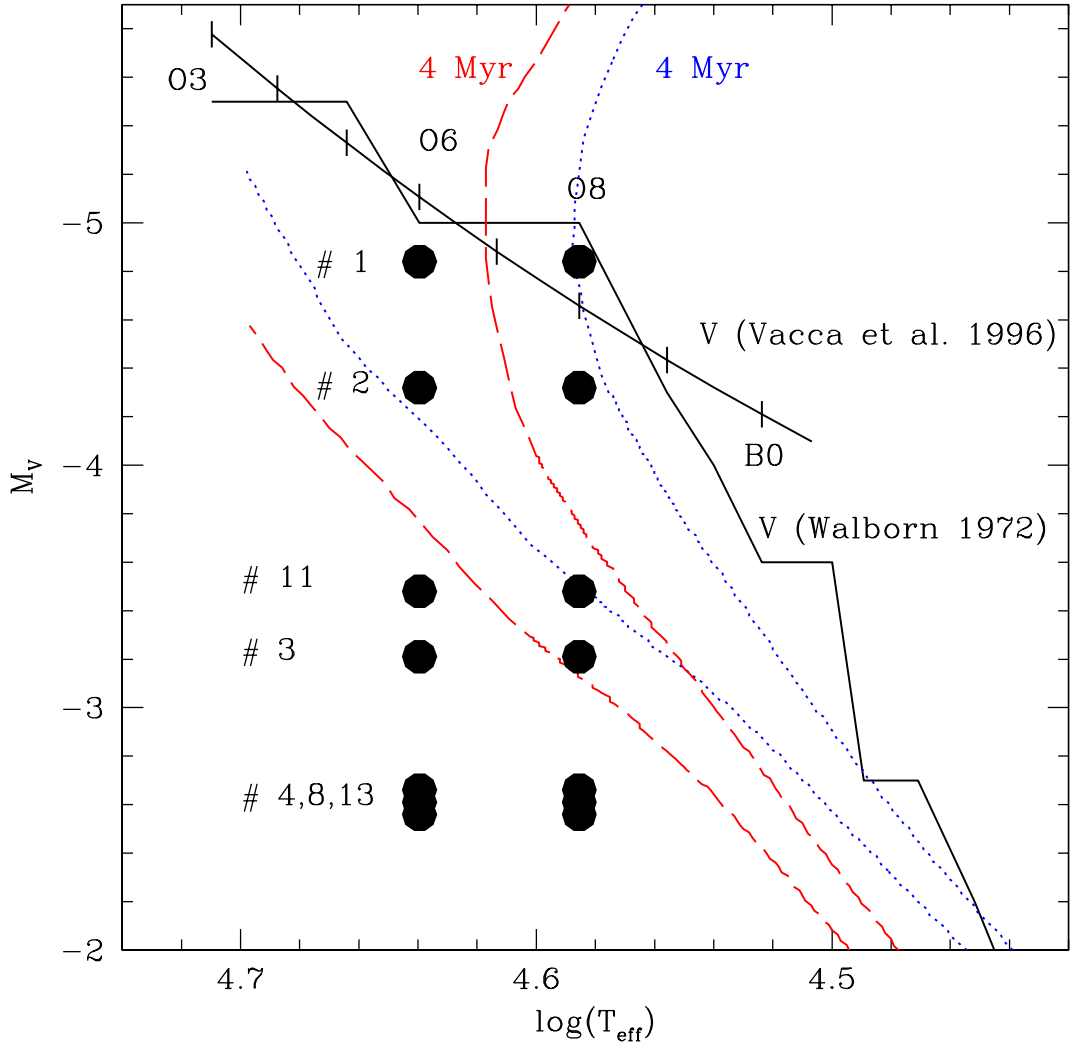}}
\caption{
The absolute $M_V$ magnitude versus effective temperature
diagram of the N81 stars is compared to the zero age main sequence
(ZAMS) and 4\,Myr isochrones at various metallicities.  The $T_{\rm
eff}$ of the stars has been estimated using an O6 or O8 spectral type
and the Vacca et al.\ (\cite{vacca}) scale. The ZAMS and 4 Myr
isochrones for a metallicity of 1/20 $Z_\odot$ (Z\,=\,0.001) are
indicated by long-dashed lines while the same pair for 1/5 $Z_\odot$
(Z\,=\,0.004) is plotted with dotted lines. The ZAMS curves are not
plotted for $ log\, (T_{\rm eff}) \geq$ 4.7 due to lack of an
appropriate conversion to $M_V$.  Also shown, as solid lines, are the
$M_V$- $T_{\rm eff}$ calibration for dwarfs from Vacca et al.\
(\cite{vacca}) as well as the one from Walborn (\cite{wal72}). Note
that four of the observed stars are situated in the HRD-like diagram
in a locus suggesting that they are either on the ZAMS or that they
have a young age. The lower luminosity stars (\# 4, 8, 13) are
possibly hotter than the ZAMS.}

\label{fig_hrd}
\end{center}
\end{figure*}

\subsection{Stellar parameters and wind properties}

To constrain the stellar wind properties we have examined the line
profiles of the strongest UV lines. The best indications for velocity
shifts come from the \civ\,\l\,1548, 1551 feature.  The profiles of
stars \#1 and \#2 show blue-shifted absorption reaching up to velocities of
 \ab\,1700 and  2000 km s$^{-1}$ respectively.  These values
are compatible with terminal velocity measurements in other SMC O
stars of similar spectral type (Walborn et al. \cite{wal95b}, Prinja
\& Crowther \cite{pc98}).  
 For star \#5 we derive a terminal velocity of 1000 km s$^{-1}$,
while in the other profiles the terminal velocities are lower.
However, these profiles  also show asymmetries which are stronger
on the red side, and  further investigations will be necessary to
understand their structure in detail. \\

Using an effective temperature derived from the estimated
spectral types (O6--O8), in particular the one from Vacca et al.\
(\cite{vacca}) for dwarfs, and the absolute magnitudes from Table 1,
we place the N81 stars (filled circles) in an HR-like diagram 
(Fig.~\ref{fig_hrd}). For further comparison we also include the mean
$M_{V}$ magnitudes for O3--B0 dwarfs based on a compilation by Vacca
et al.\ (\cite{vacca}), as well as the mean $M_V$ relation of Walborn
(\cite{wal72}) used by Walborn \& Blades (\cite{wb}) for 30 Doradus stars. 
Fig.\ \ref{fig_hrd} clearly shows that most of the stars
in N81 are sub-luminous compared to the mean $M_V$ for dwarfs.  This
finding still holds even if the spectral types are shifted by 1-2
subtypes towards later types. Compared to the so-called Vz luminosity
class, the sub-luminosity of the N81 stars is more pronounced. Our
targets are up to $\sim$ 2 mag fainter in $M_{V}$ than the mean
relation for dwarfs!  Although this is not the defining characteristic
of the Vz class, the above indications as well the weakness of the UV
wind lines further attests that these stars belong to the Vz class. \\

Two sets of ZAMS and isochrones of 4 Myr are presented in
Fig.~\ref{fig_hrd} using the Geneva stellar evolution tracks for
metallicities bracketing approximately that of the SMC (Lejeune \&
Schaerer \cite{ls}). One set has been computed for a metallicity 1/20
Solar ($Z=0.001$) and is marked with a long-dashed line, while the
second has a metallicity of 1/5 Solar ($Z=0.004$) and is marked with a
dotted line.  The observed $M_V$ and spectral types are roughly
compatible with positions close to the theoretical ZAMS or young ages.
Given mostly the lack of an accurate spectral subtype determination we
cannot firmly establish if the lower luminosity stars (\#8, \#13) are
really hotter than the ZAMS.  Atmospheric modeling is in progress to
obtain more accurate stellar parameters of these unique young stars in
the SMC (Martins et al., in preparation).  Based on the $Z\,=\,0.004$
tracks, the ZAMS luminosities and masses corresponding to the observed
$M_V$ are between $\log L/L_\odot \sim$ 4.2\,--\,5.5 and
$\sim$\,14\,--\,50 M$_\odot$ respectively.

\section{Discussion}

The {\it HST} spectra of N81 presented here are the first ones ever
obtained from a tight cluster of stars in a HEB. The reason is
that these stars, embedded in a compact emission nebula, have not been
reachable by ground-based telescopes. And even  with recent
developments in ground-based instruments, taking spectra of 
individual stars in the visible remains still practically infeasible.
As a result, contrary to other massive stars in the SMC that 
have been observed from space in
the UV (Walborn et al.\ \cite{wal95b}, \cite{wal00}, Smith Neubig \&
Bruhweiler
\cite{smith_neubig_smc}), the N81 stars lack high quality spectra in
the visible.  Our low-resolution {\it HST} spectra in the visible,
imposed by stringent time allocation constraints, were intended
to be a first exploratory such attempt. \\

Massive stars observed from the ground and also with {\it HST} are
typically much brighter than the ones seen in N81 and comparably 
they are much less affected by nebular emission and dust.
The decoupling from nebulae is presumably
due to the evolutionary state of these stars; they have had enough
time to entirely and/or effectively disrupt their \h2\, regions and
the associated dust.  
This means that those bright stars are older than the N81 members, as 
supported by their
spectra. Among the 15 SMC stars studied by Walborn et al.
(\cite{wal00}), 9 are clearly giants, 5 are peculiar and
have already developed emission line features of \niv\,\l\,4058,
\niii\,\l\,4640, or \ciii\,\l\,4650 (two of them  being on the  
main sequence),  
and  the last one is a pure main sequence. \\

The fact that all the observed exciting stars of N81 display the Vz
characteristics, further supports the very young age of the cluster
(Paper I). This observed concentration of Vz stars in a small region
is  intriguing since in the LMC region of 30 Doradus
 only 6 of the 104 O and
early B stars classified by Walborn \& Blades (\cite{wb}), that is
\ab\,6\%, belonged to the Vz category.  If we assume that all 20
bright stars we detected towards N81 (Paper I) are of O or B type,
then the lower limit for the fraction of Vz stars in N81 is nearly
35\%, which is quite considerable! \\
 
The exact nature and evolutionary stage of Vz stars is still unknown.
Presumably these objects are close to a transition from their
formation locus in the HR diagram to the main sequence.  However,
several issues regarding their properties, which are relevant to our
observations of compact low-metallicity \h2\, regions, remain open: \\

\begin{enumerate}
\item {\em Why do these stars show such weak stellar winds?}
Are their mass loss rates compatible with expectations from ``normal'' 
O stars, i.e.\ due to their reduced luminosity compared to stars with
similar effective temperatures and representative for massive stars 
in their earliest evolutionary stage?\\

\item {\em Are these objects truly on the ZAMS, blueward or 
redward of it?}  The presence or absence of massive stars on the ZAMS
-- corresponding to the locus of completely homogeneous objects
initiating H-burning -- yields information on the star formation
process.  Indeed, the apparent lack of Galactic OB stars close the
ZAMS could be due to the hiding of such stars in their parental cocoon
(Garmany et al.\ \cite{gar}) or explained by the progressive redward
bending of the upper part of the birthline due to moderate mass
accretion rates in an accretion scenario for these stars (Bernasconi
\& Maeder \cite{bern}).  In the former case the position of the
``earliest'' star visible provides thus constraints on the duration of
the hidden phase.  For the latter scenario, the position of the bluest
stars constrains the accretion rate $\dot{M}_{\rm acc}$; the existence
of massive stars close to the ZAMS requires high values of
$\dot{M}_{\rm acc}$ (Norberg \& Maeder  \cite{nm00}, Behrend
\& Maeder \cite{bm01}).\\

Given the present uncertainties on the $T_{\rm eff}$ determination of
the N81 stars, one may speculate that some of the objects are indeed
{\em hotter} than the ZAMS, as indicated by Fig. \ref{fig_hrd}.  If
true, this ``blue straggler'' like behavior could be indicative of
stellar collisions (e.g.\ Benz \& Hills \cite{bh}), and
thus a signature of formation of massive stars via this process, as
advocated among others by Bonnell et al.\ (\cite{bo}).\\

\item {\em Could some of the N81 stars still be accreting 
pre-main sequence objects?}  The stars with smaller $M_V$ magnitudes
(\# 3, 4, 8, 11, and 13) have estimated luminosities slightly above
the probable high luminosity Herbig Ae/Be pre-main sequence stars
found by Lamers et al.\ (1999) and de Wit et al.\ (2001) in the LMC
and SMC.  Their location in the HRD also coincides with the region
where the predicted birthlines follow quite closely the ZAMS.  Given the
strong indications for a very young age of N81, it is thus conceivable
that our objects are still accreting mass as part of their formation
process.  The redshifted C~{\sc iv} profiles in the above cited
objects could be an indication of ongoing accretion.  \\

\item {\em Are Vz stars related to low metallicity?}
Vz stars have also been detected in the Milky Way by Walborn and
co-workers (private communication). Therefore, it appears unlikely
that they are due to a simple metallicity effect.

\end{enumerate}

While at present these answers remain fairly speculative, our upcoming 
determinations of stellar and wind parameters and further studies of
similar objects should hopefully shed more light onto these 
issues related to the formation and early evolution of massive stars.

\section{Conclusions}

Our {\it HST}/STIS observations of the brightest massive stars
powering the high excitation compact \h2\, region SMC N81 reveal
that the stars have strikingly weak wind profiles and a pronounced
sub-luminosity which are clear indications of their early evolutionary
state. Most likely they belong to the Vz category of massive
Magellanic Cloud stars (Walborn \& Parker \cite{wp}, Walborn \& Blades
\cite{wb}), which are very young stars just arriving on the ZAMS or 
already located near it. These stars may also serve as templates
for newborn massive stars of distant metal-poor galaxies which 
cannot be individually resolved. Therefore, a more
detailed study and modeling of their properties is highly desirable as
it will help shed some more light on the intricacies and consequences
of the early stages in stellar evolution. \\

\begin{acknowledgements}   
We would like to thank the referee for a critical reading of the 
paper which contributed to its improvement. 
We are also grateful to Dr. Nolan R. Walborn (Space Telescope Science
Institute) for very helpful comments and suggestions. 
VC would like to acknowledge the financial support for this
work provided by NASA through grant number GO-8246 from the STScI,
which is operated by the Association of Universities for Research in
Astronomy, Inc., under NASA contract 26555.  DS, FM, and MH-M received
partial support from the French ``Programme National de Physique
Stellaire'' (PNPS).

\end{acknowledgements}

{}

\end{document}